\documentstyle[11pt]{article}

\begin{document}

\vspace{.7in}

\begin{center}
\Large
{\bf On the tensor field inflation in the General Relativity homogeneous cosmological model}\\
\vspace{.7in}
\normalsize
\large{ 
Poghos F. Kazarian\footnote{E-mail: pkazarian@hotmail.com}
}\\
\normalsize
\vspace{.3in}

{\em Received 19/07/2000}
\vspace{.4in}

{\em Department of Theoretical Physics, Yerevan State University, \\
1 A. Manoogian St., 375049, Yerevan, Armenia }\\
\vspace{.2in}
\end{center}
\vspace{.3in}
\baselineskip=24pt

\begin{abstract}
\noindent The homogeneous cosmological model in GR is proposed, where the vacuum energy, which can cause the inflation, is described by tensor field rather than by commonly used in inflationary scenarios scalar field. It is shown that if the initial values of the field are sufficiently big (comparable with the Planck units), under the condition of the tensor field's slow change in the beginning the regime of the quasiexponential inflation can exist. Numerical solutions for the inflationary stage are obtained that confirm the validity of the approximate solutions. Inflation takes place under wide range of initial conditions provided that the tensor field satisfies the condition imposed on the initial values of the tensor field ${\phi^0}_0(0)\approx -{\phi^i}_i(0)$ (i=1,2,3). That condition also arises from the requirement to satisfy existing observational data.

\end{abstract}


\vspace{.3in}

\noindent Since recently mostly scalar fields were used as the inflation creating force in cosmological inflationary models. Though solving the majority of problems of standard Friedman cosmology, the first models also had their owns [1-3]. The chaotic inflation [4-7] is free of these problems, but the introduction of the anthropic principle here is necessary (For detailed discussion see, e.g. [8] and the references there).As another way to avoid occurring problems in simple scenarios, models with two scalar fields responsible for the inflation [9-13] were proposed, as well as the scalar field driven inflation in non-Einsteinian theories (e.g. scalar-tensor theories [14-17]), where one scalar field is already present from the theory itself. Also we'd like to notice so-called bimetric scalar-tensor theory [18,19], which allows the extended inflationary regime, like scalar-tensor theories, but has less constraints on itself from observational data [20] (for constraints from inflation on STT, for example, see [21]). 
\noindent The use of tensor field instead of scalar fields can also be an interesting possibility. On the one hand we could have more degrees of freedom to manipulate, and on the other, these are the components of the same field bound together (let's mention that the more simple vector field obviously cannot be considered if want to satisfy the condition of the isotropy of Universe). Having more degrees of freedom brings up the requirement for the tensor field to have the initial values in the form ${\phi^i}_k \approx -b(t_0)diag(1;-1;-1;-1)$ to achieve the inflationary regime, as it will be shown below. Obviously, scalar fields have not extra degrees of freedom, and the scalar field inflation lacks such a constraint.
\noindent Let's suppose here a homogeneous inflationary model of General Relativity with the non-gravitational tensor field, introduced by the Lagrangian

\begin{equation}
L = L_0 + L_m \label{eq(1.1)}
\end{equation}

where

\begin{equation}
L_0 = \frac{1}{4}\phi_{ik;l}\phi^{ik;l} - \frac{1}{2}\phi_{ik;l}\phi^{il;k} + \frac{1}{2}{\phi^i}_{i;k}{\phi^{k;l}}_l - \frac{1}{4}{\phi^i}_{i;k}{\phi^{l;k}}_l, \label{eq(1.2)}
\end{equation}

and

\begin{equation}
L_m  = \frac{m^2}{4}( \phi_{ik}\phi^{ik} - {\phi^i}_i{\phi^k}_k ) \label{eq(1.3)}
\end{equation}

Here $\phi_{ik}$ is a tensor field, which we choose in the form

\begin{equation}
{\phi^i}_k = diag(b + \varepsilon ; b; b; b) = b(t){\delta^i}_k + \varepsilon (t){\delta^i}_0{\delta^0}_k , \label{eq2}
\end{equation}

the semicolon denotes the covariant derivative and $m$ is the tensor field mass(the light velocity everywhere is $c = 1$). As usual, let's choose the metric as [22-24]

\begin{equation}
ds^2 = dt^2 - a^2(t) [\frac{dr^2}{1 - kr^2} + r^2 (d\theta^2 + sin^2\theta d\phi^2)] , \label{eq3}
\end{equation}

where $ a(t) $ is the scale parameter or the "radius" of the Universe. Let's note that the field (4) is symmetric and satisfies the following relations, where the dot denotes the time derivative:

\begin{eqnarray}
\phi_{ik} = \phi_{ki};\nonumber
\end{eqnarray}
\begin{eqnarray}
{\phi^i}_{k;l} = {\phi^i}_{k,l} + \varepsilon ({\Gamma^i}_{0l}{\delta^0}_k - {\Gamma^0}_{kl}{\delta^0}_i);\nonumber
\end{eqnarray}
\begin{eqnarray}
{\phi^0}_{0;\alpha} = {\phi^\alpha}_{0;0} = {\phi^0}_{\alpha;0} = 0;\nonumber
\end{eqnarray}
\begin{eqnarray}
{\phi^\alpha}_{\beta;l} = {\phi^\alpha}_{\beta,l} = \dot{b}{\delta^0}_l{\delta^{\alpha}}_{\beta}; \nonumber
\end{eqnarray}
\begin{eqnarray}
{\phi^0}_{0;l} = {\phi^0}_{0,l} = (\dot{b} +\dot{\varepsilon}){\delta^0}_l;\nonumber
\end{eqnarray}
\begin{eqnarray}
{\phi^0}_{\alpha;l} = -\varepsilon{\Gamma^0}_{\alpha l}; {\phi^\alpha}_{0;l} = \varepsilon{\Gamma^\alpha}_{0l}; {\phi^0}_{\alpha;l} = -\varepsilon{\Gamma^0}_{\alpha \beta}{\delta^\beta}_l ; {\phi^\alpha}_{0;l} = \varepsilon{\Gamma^\alpha}_{0\beta}{\delta^\beta}_l ;\nonumber
\end{eqnarray}
\begin{eqnarray}
{\phi^i}_i = 4b + \varepsilon ; \phi^{ik} \phi_{ik} = {\phi^i}_k {\phi^k}_i = (b + \varepsilon)^2 +3b^2 \label{eq4}
\end{eqnarray}	

In (4) the Latin indices run from 0 to 4 and Greek indices - from 0 to 3. Considering (6), (3) can be rewritten as

\begin{equation}
L_m = \frac{m^2}{4}( [(b + \varepsilon)^2 + 3b^2] - (4b + \varepsilon)^2 ) = -\frac{3}{2}m^2 b (2b + \varepsilon) \label{eq5}
\end{equation}

Considering that

\begin {eqnarray}
\phi_{ik;l}\phi^{ik;l} = [( \dot{b} + \dot{\varepsilon} )^2 + 3{\dot{b}}^2] - 2{\varepsilon}^2g^{\mu\sigma}{\Gamma^\alpha}_{0\gamma}{\Gamma^0}_{\alpha\sigma} ;\nonumber
\end{eqnarray}
\begin{eqnarray}
\phi_{ik;l}\phi^{il;k} = ( \dot{b} + \dot{\varepsilon} )^2  -  {\varepsilon}^2g^{\alpha\beta}{\Gamma^0}_{\alpha\gamma}{\Gamma^\gamma}_{0\beta} + 2\varepsilon b \dot{ln}\sqrt{-g} ;\nonumber
\end{eqnarray}
\begin{eqnarray}
{\phi^i}_{i;k}{\phi^k;l}_l = (4\dot{b} + \dot{\varepsilon)})[( \dot{b} + \dot{\varepsilon} )^2  -  {\varepsilon}g^{\alpha\beta}{\Gamma^0}_{\alpha\beta}] ; {\phi^i}_{i;k}{\phi^l;k}_l = (4\dot{b} + \dot{\varepsilon)})^2 \label{eq6}
\end {eqnarray}

for $L_0$, determined by (2), we obtain

\begin{eqnarray}
L_0 = \frac{3}{2}{\dot{b}}^2 + \frac{1}{2}\varepsilon(2\dot{b} + \dot{\varepsilon})\dot{ln}\sqrt{-g} ;\nonumber
\end{eqnarray}
\begin{eqnarray}
\dot{ln}\sqrt{-g} = 3\frac{\dot{a}}{a} \equiv 3H \label{eq7}
\end{eqnarray}

where, as usual, $H$ denotes the Hubble parameter. Finally, the total action can be written as

\begin{equation}
S_{tot} = \frac {3}{2} \{ \int \sqrt{-g}d\Omega [\frac {1}{4\pi G}( \frac{\ddot{a}}{a} + {\frac{\dot{a}}{a}}^2 + \frac{k}{a^2} ) + \frac{\dot{a}}{a}\varepsilon(2\dot{b} + \dot{\varepsilon}) - {\dot{b}}^2 - m^2 b (2b+ \varepsilon)]\} \label{eq8}
\end{equation}

where $G$ is the Newton constant, $k = -1;0;1$ for open, flat and closed Universe and  $\sqrt{-g} = \frac{a^3(t)r^2 sin\theta}{\sqrt{1-kr^2}}$. The variation of (10) will give us the field equations or the equations of motion. Varying (1) by $\varepsilon$ , $b$ , and scale factor $a$ will provide us with the field equations:

\begin{eqnarray}
2\dot{b}H = \varepsilon (\dot{H} + 3H^2) + m^2 b;\nonumber
\end{eqnarray}
\begin{eqnarray}
\dot{\varepsilon}H + \dot {H}\varepsilon - \ddot{b} + 3H( \varepsilon H  - \dot{b}) + m^2(2b + \varepsilon/2)= 0;\nonumber
\end{eqnarray}
\begin{eqnarray}
\frac{1}{4\pi G}(2\dot{H} + 3H^2 + \frac{k}{a^2})  -{\{ \varepsilon(2\dot{b} + \dot{\varepsilon}) \}}^{.} - {\dot{b}}^2 - 3m^2 b (2b+ \varepsilon) = 0 \label{eq9}
\end{eqnarray}

Using the first equation of (11), the second one can be also rewritten as

\begin{equation}
\ddot{b}  = H (\dot{\varepsilon} -  \dot{b}) + m^2 (b + \varepsilon/2) \label{eq10}
\end{equation}

Assuming [4,25] that in the beginning (at times comparable with the Planck time $t_p$ ) the tensor field was ${\phi^i}_k \sim M_p$ (where $M_p$ is the Planck mass) and was slowly changing, for we can ignore the second derivatives (${{\ddot{\phi}}^i}_k , ({{\dot{\phi}}^i}_k)^2 \ll V({\phi^i}_k) $) , and also considering that the scale factor $a$ is sufficiently big (the term $\frac{k}{a^2}$ can be ignored ), from (12) we obtain

\begin{equation}
\dot{H} = \frac{m^2}{2} \frac{2+\dot{\varepsilon}/\dot{b}}{1 - \dot{\varepsilon}/\dot{b}}, \label{eq11}
\end{equation}

and, using field equations,

\begin{equation}
H^2 = m^2( 4\pi Gb(2b+\varepsilon) +  \frac{1}{3}\frac{2+\dot{\varepsilon}/\dot{b}}{\dot{\varepsilon}/\dot{b} - 1}) \label{eq12}
\end{equation}

From here it's obvious, that $\dot{H}\ll H^2$ (for the change of $b$ is comparable to the one of $\varepsilon$ at $b, \varepsilon \sim M_p$ ) and for slowly changing matter field (${{\dot{\phi}}^i}_k / {\phi^i}_k$) in the early stages of the existence of Universe $H \approx const$ . In this approximation the field equations (11) become

\begin{eqnarray}
2\dot{\varepsilon}H  = 3H^2 \varepsilon - m^2( b + \varepsilon ) ;\nonumber
\end{eqnarray}
\begin{eqnarray}
2\dot{b}H = 3H^2 \varepsilon  + m^2 b ;\nonumber
\end{eqnarray}
\begin{eqnarray}
\frac{1}{4\pi G} H^2 = m^2 b (2b+ \varepsilon) \approx \frac{2}{3} V({\phi^i}_k) \label{eq13}
\end{eqnarray}

Let's study the behavior of $b$ and $\varepsilon$ . Introducing new variables $A = \varepsilon / b , B = \dot{b} / b$ , from (15) follows

\begin{eqnarray}
\dot{A} + AB - B + E( 2 + A ) = 0 ;\nonumber
\end{eqnarray}
\begin{eqnarray}
B - E = \frac{3}{2} HA ;\nonumber
\end{eqnarray}
\begin{eqnarray}
E \equiv \frac{m^2}{2H} = const \label{eq14}
\end{eqnarray}

After integration

\begin{equation}
t = \frac{1}{A_1 - A_2} ln |{\frac{A - A_2}{A - A_1}}| , \label{eq15}
\end{equation}

where

\begin{equation}
A_{1,2} =(\mu + \frac{1}{2}) \pm \sqrt{(\mu +\frac{1}{2})^2 -\mu} \label{eq15}
\end{equation}

Here $\mu \equiv \frac{m^2}{3H^2} \geq 0$ . Finally,

\begin{equation}
A(t) = \frac{A_2 - A_1 C_0 exp[3H\sqrt{(\mu + \frac{1}{2})^2- \mu}t]}{1 - C_0 exp[3H\sqrt{(\mu + \frac{1}{2})^2-\mu}t]} , \label{eq16}
\end{equation}

and $C_0 = \frac{A(0) - A_2}{A(0) - A_1}$. In the initial period $A \approx const$ , and from (16)

\begin{eqnarray}
b \approx b_0 exp(\frac{3}{2}H(A + \mu)t) \leq b_0 ;\nonumber
\end{eqnarray}
\begin{eqnarray}
\varepsilon \approx Ab; A \approx const \label{eq17}
\end{eqnarray}

Because
\begin{eqnarray}
b \approx b_0 + ( [3A + \frac{m^2}{H^2}\frac{Ht}{2}]) ;\nonumber
\end{eqnarray}
\begin{eqnarray}
(V({\phi^i}_k)) = V(b) \approx m^2 {b_0}^2 (-A - 2) ; \label{eq18}
\end{eqnarray}

then for the scale factor $a$ we'll obtain

\begin{equation}
a \approx a_0 exp ( \frac{6\pi}{{M_p}^2}[{b_0}^2 - b^2 (t)](-A - 2) ) \label{eq19}
\end{equation}

If the term $(-A - 2)$ in (20) is positive, then we have the (quasi)exponential growth of the scale factor in this model, which leads to the requirement $A\leq -2$. Let's also note that the full expansion of Universe under these conditions will be

\begin{equation}
P \approx exp ( \frac{3{b_0}^2 (-A - 2)}{{M_p}^2}) \sim exp ( \frac{\pi \sqrt{2}{M_p}^2}{3m^2} ) , \label{eq20}
\end{equation}

considering that $\varepsilon_0 , b_0 \sim M_p$ , the initial values of the tensor field components are approximately of the Planck mass.
\noindent We've shown that under certain conditions there is a possibility of quasiexponential inflation in the Universe with the vacuum energy defined by a tensor field (1)-(3). However, questions still remain about the realization of these conditions in the early Universe, like for how long the approximations of the proposed model can describe with sufficient precision the real evolution of the Universe and for how long this inflationary regime will last. 
\noindent The numerical solutions of the field equations (11) show that the approximation ( $H \approx const$, ${{\dot{\phi}}^i}_k / {\phi^i}_k \approx const$, and ${{\ddot{\phi}}^i}_k , ({{\dot{\phi}}^i}_k)^2 \ll V({\phi^i}_k) $) is valid through the inflationary stage. The requirements necessary for the inflationary regime realization are

\begin{eqnarray}
V_0({\phi^i}_k) \leq 1 ;\nonumber
\end{eqnarray}
\begin{eqnarray}
|2b_0 + \varepsilon_0| \ll |b_0| , |\varepsilon_0| ;\nonumber 
\end{eqnarray}
\begin{eqnarray}
0 < H_0 \ll |b_0| , |\varepsilon_0| ; \label{eq21}
\end{eqnarray}

(here and below $c \equiv 1$ , $G = {M_p}^{-2} \equiv 1$). As far as $\dot{b} , \dot{\varepsilon}$ satisfy (24), we have considerable freedom of choice of their exact initial values, for that does not affect noticeably the numerical solutions because in the negligible time after the beginning $ |b_0| , |\varepsilon_0| = const \ll 1$ and remain so during the inflation. The choice of $H_0$ also does not affect qualitatively the solutions, although quantitatively $t_{infl}$ (the time of increase of $a(t)$ in $\approx10^{30}$ times) is somewhat shorter for greater $H_0$. Otherwise, regardless to the initial value, $H$ rapidly becomes (and stays during the inflation) constant: $H = const \ll 1$. The decrease (increase) of the tensor field mass causes the increase (decrease) of $t_{infl}$. Also, for a greater mass the greater $b_0 , \varepsilon_0$ are required to satisfy (24) and achieve the satisfactory inflationary regime.
\noindent It seems that the only strict requirement for achieving inflation is $|2b_0 + \varepsilon_0| \ll |b_0| , |\varepsilon_0| $ (or, equivalently, $\varepsilon_0  \approx -2b_0$ ).Any noticeable deviation from that condition makes the inflationary regime impossible. Because $b \approx b_0 = const ,  \varepsilon \approx \varepsilon_0 = const$ during the inflation, $|2b_0 + \varepsilon_0|$ also stays at its initial value. The behavior of scale parameter $a(t)$ doesn't differ qualitatively  for different sets of parameters. It increases about $10^{30}$ times during $t_{infl}$. The growth of $a(t)$ becomes somewhat faster (though of the same order of magnitude) with the increase of the mass and initial values of the tensor field components and the decrease of $|2b_0 + \varepsilon_0|$. Values of parameters for different initial conditions are presented in the Tables 1-3.
\noindent Observations impose several constraints on initial values and field parameters (see, e.g. [8] and the references there). For instance, the requirement of having the inflationary regime for sufficiently long time ($t_{infl} \sim M_p$)imposes the upper limits on the field mass ($m \leq 10^{-6}$). Also, the density fluctuations should be not greater than $\frac{\delta\varrho}{\varrho} = 10^{-4} - 10^{-5}$, or

\begin{equation}
\frac{\delta\varrho}{\varrho} = \frac{1}{ V({\phi^i}_k)} \frac{dV}{d{\phi^i}_k} \delta {\phi^i}_k \approx \frac{\delta (2b + \varepsilon)}{2b + \varepsilon} \approx \frac{H}{2\pi(2b + \varepsilon)} = 10^{-4} - 10^{-5} ; \label{eq22}
\end{equation}

or $2b + \varepsilon \sim \frac{H}{2\pi}(10^{-4} - 10^{-5})$ (considering $|2b + \varepsilon| \ll |b|$). So (25) restricts the value of $(2b + \varepsilon)$ further for a given $H$, which, in its turn, restricts the values of $b$ and $\varepsilon$ (therefore, the values of $b_0$ and $\varepsilon_0$, for $b , \varepsilon \approx const$) to satisfy (24) (see the table).
\noindent Numerical simulations show that for a broad range of the tensor field parameters defined by (24), (25) the proposed model allows the existence of the inflationary stage of the expansion of the Universe, and those solutions are stable for broad range of initial values (satisfying (24), (25)). Numerical solutions also confirm the predictions of the approximate solutions (19)-(22) and show that the approximation is valid during the inflation. From the comparison of the numerical and approximate solutions it also follows that the parameter $A$ in (19)-(22)is constant during the inflation and 

\begin{equation}
A \leq -2 \label{eq23}
\end{equation}

for achieving the inflationary regime in the model. Let's note that the condition $ \varepsilon_0 \approx -2b_0 $ imposed on the initial values of the field is the requirement that inflation in the model can be achieved if the field is chosen initially in the form of ${\phi^i}_k(0) \approx diag(-b;b;b;b)$,  where $b \equiv b(t) \approx b_0$ during the inflation.

\begin{table}[h]
\begin{tabular}{|c|c|c|c|c|c|c|c|}
\hline
$m$& $H_0   $& $H(t_{infl})    $& $  |b| \approx |b_0|  $& $  |\varepsilon| \approx |\varepsilon_0| $& $  |2b + \varepsilon|  $& $  t_{infl}  $& $  a(t_{infl})  $\\
\hline
$10^{-6}$& $\sim 10^8$& $\sim 4*10^{-7}$& $\leq 10^{15}$& $\leq 2*10^{15}$& $\leq 10^{-2} - 10^{-3}$& $\sim 1.44*10^8$& $\sim 8.9*10^{29}$\\
\hline
$10^{-6}$& $\sim 10^{-8}$& $\sim 4*10^{-7}$& $\leq 10^{15}$& $\leq 2*10^{15}$& $\leq 10^{-2} - 10^{-3}$& $\sim 1.7*10^8$& $\sim 9.9*10^{29}$\\
\hline
$10^{-7}$& $\sim 10^8$& $\sim 4*10^{-8}$& $\leq 10^{17}$& $\leq 2*10^{17}$& $\leq 10^{-3} - 10^{-4}$& $\sim 1.41*10^9$& $\sim 8.7*10^{29}$\\
\hline
$10^{-7}$& $\sim 10^{-9}$& $\sim 4*10^{-8}$& $\leq 10^{17}$& $\leq 2*10^{17}$& $\leq 10^{-3} - 10^{-4}$& $\sim 1.41*10^9$& $\sim 9.9*10^{29}$\\
\hline
$10^{-8}$& $\sim 10^6$& $\sim 4*10^{-9}$& $\leq 10^{20}$& $\leq 2*10^{20}$& $\leq 10^{-4} - 10^{-5}$& $\sim 1.44*10^{10}$& $\sim 8.9*10^{29}$\\
\hline
$10^{-8}$& $\sim 10^{-10}$& $\sim 4*10^{-9}$& $\leq 10^{20}$& $\leq 2*10^{20}$& $\leq 10^{-4} - 10^{-5}$& $\sim 1.7*10^{10}$& $\sim 9.9*10^{29}$\\
\hline
$10^{-9}$& $\sim 10^5$& $\sim 4*10^{-10}$& $\leq 10^{23}$& $\leq 2*10^{23}$& $\leq 10^{-5} - 10^{-6}$& $\sim 1.43*10^{11}$& $\sim 8.4*10^{29}$\\
\hline
$10^{-9}$& $\sim 10^{-11}$& $\sim 4*10^{-10}$& $\leq 10^{23}$& $\leq 2*10^{23}$& $\leq 10^{-5} - 10^{-6}$& $\sim 1.7*10^{11}$& $\sim 9.5*10^{29}$\\
\hline
\end {tabular}
\caption{ Values of the inflationary model's parameters for different initial conditions}
\end{table}

\begin{table}[h]
\begin{tabular}{|c|c|c|c|c|c|c|c|c|c|c|c|}
\hline
$t$& $1.00*10^{0} $& $1.00*10^{6}$& $1.00*10^{7} $& $2.70*10^{7} $& $5.70*10^{7} $& $1.14*10^{8} $& $1.42*10^{8} $& $1.70*10^{8} $\\
\hline
$H$& $0.5*10^{-8}$& $3.45*10^{-7}$& $4.08*10^{-7}$& $4.08*10^{-7}$& $4.08*10^{-7}$& $4.08*10^{-7}$& $4.08*10^{-7}$& $4.08*10^{-7}$\\
\hline
$a$& $1$& $1.23$& $47.3$& $4.9*10^{4}$& $1.02*10^{10}$& $1.3*10^{20}$& $1.2*10^{25}$& $9.9*10^{29}$\\
\hline
\end {tabular}
\caption{Typical behavior of the scale parameter $a(t)$ and the Hubble parameter $H$ for $H_0\ll 1$ ($m = 10^{-6}, M_p = 1)$}
\end{table}

\begin{table}[h]
\begin{tabular}{|c|c|c|c|c|c|c|c|c|c|c|}
\hline
$t$& $1.00*10^{0} $& $1.00*10^{6} $& $3.00*10^{6} $& $6.00*10^{6} $& $3.10*10^{7} $& $8.70*10^{7} $& $1.16*10^{8} $& $1.44*10^{8} $\\
\hline
$H$& $0.5*10^{8}$& $4.85*10^{-7}$& $4.09*10^{-7}$& $4.08*10^{-7}$& $4.08*10^{-7}$& $4.08*10^{-7}$& $4.08*10^{-7}$& $4.08*10^{-7}$\\
\hline
$a$& $1$& $5.75*10^{4}$& $1.34*10^{5}$& $4.56*10^{5}$& $1.24*10^{10}$& $1.05*10^{20}$& $1.46*10^{25}$& $8.95*10^{29}$\\
\hline
\end {tabular}
\caption{Typical behavior of the scale parameter $a(t)$ and the Hubble parameter $H$ for $H_0\geq 1$ ($m = 10^{-6}, M_p = 1)$}
\end{table}

\vspace{.3in}
{\bf References}
\vspace{.3in}

\begin{tabular}{ll}
1. &  A.H. Guth, Phys. Rev. D 23, 347 (1981)\\
2. &  A.D. Linde, Phys. Lett. 108B, 389 (1982)\\
3. &  A. Albrecht, P.J. Steinhardt, Phys. Rev. Lett. 48, 1220 (1982)\\
4. &  A.D. Linde, Phys. Lett. 129B, 177 (1983)\\
5. &   A.D. Linde, Mod. Phys. Lett. A 1, 81 (1986)\\
6. &   A.D. Linde, Phys. Lett. 175, 395 (1986)\\
7. &   A.D. Linde, Phys. Scripta T15, 169 (1987)\\
8. &   A.D. Linde, Particle Physics and Inflationary Cosmology,Harwood,Chur,Switzerland(1990)\\
9. &   J. Garcia-Bellido, D. Wands, Phys. Rev. D 54, 7181 (1996); astro-ph/9608042\\
10. & L.A. Kofman, A.D. Linde, Nucl. Phys. B 282, 555 (1987)\\
11. & A.D. Linde, Phys. Rev. D 49, 748 (1994)\\
12. & J. Garcia-Bellido, A.D. Linde, Phys. Rev. D 55, 7480 (1997)\\
13. & J.S. Silk, M.S. Turner, Phys. Rev. D 35, 419 (1987)\\
14. & D. La, D.J. Steinhardt, Phys. Rev. Lett. 62, 376 (1989)\\
15. & E.J. Weinberg, Phys. Rev. D 40, 3950 (1989)\\
16. & J.D. Barrow, K. Maeda, Nucl. Phys. B341, 294 (1990)\\
17. & F.S. Accetta, J.J. Trester, Phys. Rev. D 39, 2854 (1989)\\
18. & L.Sh. Grigorian, A.A. Saharian, Astrophysics 31, 359 (1989)\\
19. &  L.Sh. Grigorian, A.A. Saharian, Astrophys. Space Sci. 167, 271 (1990)\\
20. &  A.A. Saharian, Astrophysics 37, 261 (1994)\\
21. & J. Garcia-Bellido, D. Wands, Phys. Rev. D 53, 5437 (1996); gr-qc/9506050\\
22. &  A. Friedmann, Z. Phys. 10, 377 (1922)\\
23. & H.P. Robertson, Rev. Mod. Phys. 5, 62 (1933)\\
24. &  A.G. Walker, J. Lond. Math. Soc. 19, 219 (1944)\\
25. &  A.D. Linde, Phys. Lett. 162B, 281 (1985); Suppl. Progr. Theor. Phys. 85, 279 (1985)
\end{tabular}

\end{document}